\documentclass{PoS}

\usepackage{amsmath}
\usepackage{amssymb}

\title{Heavy flavour production in the SACOT-$m_{\rm T}$ scheme}

\ShortTitle{Heavy flavour production in the SACOT-$m_{\rm T}$ scheme}

\author{Ilkka Helenius \\
        University of Jyvaskyla, Department of Physics, P.O. Box 35, FI-40014 University of Jyvaskyla, Finland \\
        Institute for Theoretical Physics, T\"ubingen University, Auf der Morgenstelle 14, 72076 T\"ubingen, Germany \\
        E-mail: \email{ilkka.m.helenius@jyu.fi}}

\author{\speaker{Hannu Paukkunen}\\
        University of Jyvaskyla, Department of Physics, P.O. Box 35, FI-40014 University of Jyvaskyla, Finland \\
        Helsinki Institute of Physics, P.O. Box 64, FI-00014 University of Helsinki, Finland \\
        E-mail: \email{hannu.paukkunen@jyu.fi}}

\abstract{
The hadroproduction of heavy-flavoured mesons has recently attracted a growing interest e.g. within the people involved in global analysis of proton and nuclear parton distribution functions, saturation physics, and physics of cosmic rays. In particular, the D- and B-meson measurements of LHCb at forward direction are sensitive to gluon dynamics at small $x$ and are one of the few perturbative small-$x$ probes before the next generation deep-inelastic-scattering experiments. In this talk, we will concentrate on the collinear-factorization approach to inclusive D-meson production and describe a novel implementation --- SACOT-$m_{\rm T}$ --- of the general-mass variable flavour number scheme (GM-VFNS). In the GM-VFNS framework the cross sections retain the full heavy-quark mass dependence at $p_{\rm T}=0$, but gradually reduce to the ordinary zero-mass results towards asymptotically high $p_{\rm T}$. However, the region of small (but non-zero) $p_{\rm T}$ has been somewhat problematic in the previous implementations of GM-VFNS, leading to divergent cross sections towards $p_{\rm T} \rightarrow 0$, unless the QCD scales are set in a particular way. Here, we provide a solution to this problem. In essence, the idea is to consistently account for the underlying energy-momentum conservation in the presence of a final-state heavy quark-antiquark pair. This automatically leads to a well-behaved GM-VFNS description of the cross sections across all $p_{\rm T}$ without a need to fine tune the QCD scales. The results are compared with the LHCb data and a very good agreement is found. We also compare to fixed-order based calculations and explain why they lead to approximately a factor of two lower D-meson production cross sections than the GM-VFNS approach.
}

\FullConference{International Conference on Hard and Electromagnetic Probes of High-Energy Nuclear Collisions\\
		30 September - 5 October 2018\\
		Aix-Les-Bains, Savoie, France}

\begin{document}

\section{Motivation}

The potential of D- and B-meson production as a constraint for parton distributions (PDFs) has been recently under active investigation \cite{Zenaiev:2015rfa,Gauld:2016kpd,Kusina:2017gkz}. The heavy-quark mass provides a hard scale offering a possibility to use perturbative QCD for production of heavy-flavoured mesons even down to zero transverse momentum, $P_{\rm T} = 0$. While the general-purpose PDFs commonly used for LHC phenomenology are defined in general-mass variable flavour number schemes (GM-VFNS) \cite{Thorne:2008xf}, there are no public GM-VFNS tools for heavy-flavoured meson hadroproduction available. This was the motivation for our study \cite{Helenius:2018uul} which we summarize here.

\section{Heavy-flavour production in fixed flavour-number schemes}

In fixed flavour-number schemes (FFNS), the heavy quarks $Q$ are produced in three partonic processes $g+g \rightarrow Q + X, \, q+\overline{q} \rightarrow Q + X, \, q+g \rightarrow Q + X\,$. The rapidity- ($y$) and transverse-momentum ($p_{\rm T}$) differentiated cross section for producing heavy quarks can be written as a convolution of PDFs $f_i^{h}(x_1,\mu^2_{\rm fact})$ and partonic cross sections $d\hat\sigma$ as
\begin{align}
\frac{d\sigma(h_1 + h_2 \rightarrow {\rm Q} + X)}{dp_{\rm T}dy} = & \sum _{ij}
\int dx_1 dx_2 
         { f_i^{h_1}(x_1,\mu^2_{\rm fact}) }
            { \frac{d\hat{\sigma}^{ij\rightarrow {\rm Q} + X}(\tau_1, \tau_2, m^2, \mu^2_{\rm ren}, \mu^2_{\rm fact})}{dp_{\rm T} dy} }
         { f_j^{h_2}(x_2,\mu^2_{\rm fact}) }  \,,           
        \nonumber
\end{align}
where 
$\tau_1 \equiv {p_1 \cdot p_3}/{p_1 \cdot p_2} = {m_{\rm T}e^{-y}}/({\sqrt{s} x_2})$, \, 
$\tau_2 \equiv {p_2 \cdot p_3}/{p_1 \cdot p_2} = {m_{\rm T}e^{y}}/({\sqrt{s} x_1})$, and $m_{\rm T}$ represents the transverse mass $m^2_{\rm T} = {p_{\rm T}^2+m^2}$. Here $p_{1,2}$ refer to the momenta of the incoming partons, $p_3$ is the momentum of the outgoing heavy quark $Q$, and $m$ denotes the heavy-quark mass. The renormalization and factorization scales are denoted by $\mu^2_{\rm ren}$ and $\mu^2_{\rm fact}$. At high $p_{\rm T}$ the FFNS cross section diverges logarithmically $d\sigma \sim \log(p_{\rm T}^2/m^2)$, so the framework is reliable only at low $p_{\rm T}$.

To convert the parton-level cross sections to hadronic ones, the partonic spectrum is typically folded with a $Q\rightarrow h_3$ fragmentation functions (FFs) ${D_{Q \rightarrow h_3}(z) }$, fitted to e$^+$e$^-$ data. For this we must define a fragmentation variable $z$ which is, however, ambiguous in the presence of massive partons and hadrons. As a working assumption, we shall define $z$ as the fraction of fragmenting heavy-quark's energy carried by the outgoing hadron $h_3$ in the hadronic center-of-mass frame, $z \equiv {E_{\rm hadron}}/{E_{\rm parton}}$. Together with the assumption of collinear fragmentation, this leads to	
\begin{align}
\frac{d\sigma(h_1 + h_2 \rightarrow h_3 + X)}{dP_{\rm T}dY} = & \sum _{ij}
\int \frac{dz}{z} dx_1 dx_2
            { f_i^{h_1}(x_1,\mu^2_{\rm fact}) }
            {\frac{d\hat{\sigma}^{ij\rightarrow {\rm Q} + X}}{dp_{\rm T} dy} }
            {f_j^{h_2}(x_2,\mu^2_{\rm fact})}             
            {D_{Q \rightarrow h_3}(z) }             
            \nonumber
\end{align}
where the partonic (lower case) and hadronic variables (upper case) are related as
\begin{align}
p_{\rm T}^2 & = \frac{M^2_{\rm T}\cosh^2 Y - z^2m^2}{z^2} \left(1 + \frac{M_{\rm T}^2\sinh^2 Y}{P^2_{\rm T}}\right)^{-1} \xrightarrow{P_{\rm T} \rightarrow \infty} \left(\frac{P_{\rm T}}{z}\right)^2 \nonumber \,  \\ 
    y & = \sinh^{-1} \left(\frac{M_{\rm T} \sinh Y}{P_{\rm T}} \frac{p_{\rm T}}{m_{\rm T}}\right) \xrightarrow{P_{\rm T} \rightarrow \infty} Y \,  \nonumber
\end{align}
where $M_{\rm T} = \sqrt{P_{\rm T}^2+M_{h_3}^2}$ is the hadronic transverse mass. A framework very similar to this has been compared with the LHCb data e.g. in Ref.~\cite{Gauld:2015yia}, and the typical situation is that the calculations undershoot the data by a factor of two or so, though within the large scale uncertainties there is still a fair agreement.

\section{From FFNS to GM-VFNS heuristically}

\begin{figure}[htb!]
\centering
\includegraphics[width=0.90\linewidth]{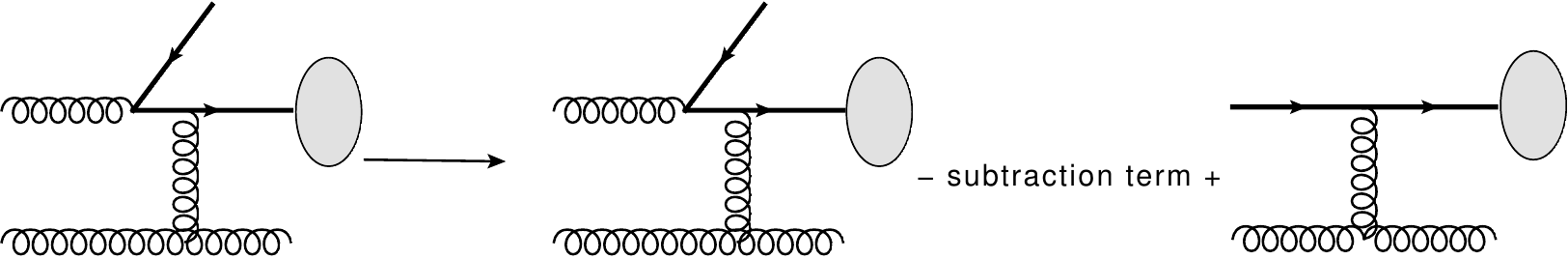}
\\
\hspace{-0.3cm} (a) \hspace{4.5cm} (a) \hspace{2.5cm} (b) \hspace{2.0cm} (c)
\caption{A schematic representation of how to deal with the initial-state logarithms.}
\label{fig:ini}
\end{figure}

The GM-VFNS framework can be derived from FFNS by resumming the $\log(p_{\rm T}^2/m^2)$ terms present in the FFNS partonic cross sections. The diagram (a) in Figure~\ref{fig:ini} shows an NLO diagram in which an incoming gluon splits into a $Q\overline{Q}$ pair giving rise to a collinear logarithm $\sim \log(p_{\rm T}^2/m^2)$. This is just the first term of the whole tower of terms that are in variable flavour number scheme resummed into the heavy-quark PDF $f_Q^{h_1}$.
This resummation can be effectively done by including the heavy-quark initiated contribution (c) and a term (b) that subtracts the overlap between diagrams (a) and (c). We may write the contribution from the $Qg\rightarrow Q+X$ channel as 
\begin{align}
\int \frac{dz}{z}dx_1 dx_2 \,
            { f_Q^{h_1}(x_1,\mu^2_{\rm fact}) }
            {\frac{d\hat{\sigma}^{Qg\rightarrow  {\rm Q} + X}(\tau_1, \tau_2)}{dp_{\rm T} dy} }
            { f_g^{h_2}(x_2,\mu^2_{\rm fact}) }             
            {D_{Q \rightarrow h_3}(z)}
        \nonumber \,.
\end{align}
The compensating subtraction term is obtained from the above expression by swapping the heavy-quark PDF with its perturbative expression to first order in $\alpha_s$,
\begin{equation}
f_Q(x,\mu^2_{\rm fact}) = \left(\frac{\alpha_s}{2\pi}\right) \log\left(\frac{\mu^2_{\rm fact}}{m^2}\right) \int_x^1 \frac{\mathrm{d}\ell}{\ell} P_{qg}\left(\frac{x}{\ell}\right) f_g(\ell,\mu^2_{\rm fact}) \,, \nonumber
\end{equation}
where $P_{qg}$ is the standard gluon-to-quark splitting function. As is well known \cite{Thorne:2008xf}, the GM-VFNS framework contains an inherent scheme dependence which leaves us with some freedom to choose the exact form of $d\hat{\sigma}^{Qg\rightarrow  {\rm Q} + X}(\tau_1, \tau_2)$ in the above expressions. In practice, the only requirement is that we must recover the zero-mass expressions at high $p_{\rm T}$,
\begin{align}
            {\frac{d\hat{\sigma}^{Qg\rightarrow  {\rm Q} + X}(\tau_1, \tau_2)}{dp_{\rm T} dy}}
        \xrightarrow{p_{\rm T} \rightarrow \infty}
            {\frac{d\hat{\sigma}^{qg\rightarrow  {\rm q} + X}(\tau_1, \tau_2)}{dp_{\rm T} dy}}
        \,\,\,\, ({\rm q = light \ quark}) \,.
\nonumber
\end{align}
The simplest option is clearly to use the zero-mass expressions from the outset, ${d\hat{\sigma}^{Qg\rightarrow  {\rm Q} + X}(\tau_1, \tau_2)} \equiv {d\hat{\sigma}^{qg\rightarrow  {\rm q} + X}(\tau_1, \tau_2)}$ and also to forget completely about the heavy-quark mass in the kinematics, $\tau_{1,2} \rightarrow {p_{\rm T}e^{\mp y}}/({\sqrt{s} x_{2,1}})$. This defines the so-called SACOT scheme \cite{Kniehl:2004fy}. The problem of this scheme is that since the partonic cross sections behave as ${d\hat{\sigma}^{qg\rightarrow {\rm q} + X}}/d^3p \xrightarrow{p_{\rm T} \rightarrow 0} \left(\tau_{1,2}\right)^{-n}$, it leads to infinite (positive or negative) production cross sections towards $P_{\rm T} \rightarrow 0$. This unphysical behaviour can be neatly avoided in what we call here the SACOT-$m_{\rm T}$ scheme \cite{Helenius:2018uul}: The idea is to retain the $Q\overline{Q}$-pair kinematics also for the $Qg\rightarrow Q+X$ channel, implicitly understanding that the final state must still contain the $\overline{Q}$. With this physical motivation, we define ${d\hat{\sigma}^{Qg\rightarrow  {\rm Q} + X}(\tau_1, \tau_2)} \equiv {d\hat{\sigma}^{qg\rightarrow  {\rm q} + X}(\tau_1, \tau_2)}$ taking $\tau_{1,2} = {m_{\rm T}e^{\mp y}}/({\sqrt{s} x_{2,1}})$ as in the massive FFNS case. This automatically leads to finite cross sections in the $P_{\rm T} \rightarrow 0$ limit.

\begin{figure}[htb!]
\centering
\includegraphics[width=0.80\linewidth]{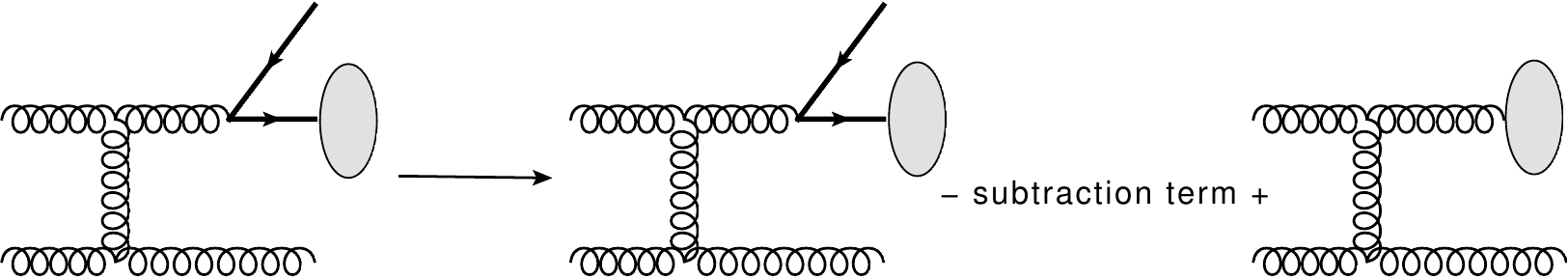}
\caption{A schematic representation of how to deal with the final-state logarithms.}
\label{fig:fin}
\end{figure}

There are also collinear logarithms coming from the final-state e.g. when --- as in Figure~\ref{fig:fin} above --- an outgoing gluon splits into a $Q\overline{Q}$ pair. 
In this case the $\log(p_{\rm T}^2/m^2)$ terms are resummed into the scale-dependent gluon FFs, $D_{g \rightarrow h_3}(z,\mu^2_{\rm frag})$. Thus, in GM-VFNS one has also the contribution of the $gg\rightarrow g+X$ channel,
\begin{align}
\int \frac{dz}{z}dx_1 dx_2 \, 
            { f_g^{h_1}(x_1,\mu^2_{\rm fact}) }
            {\frac{d\hat{\sigma}^{gg\rightarrow  g + X}(\tau_1, \tau_2)}{dp_{\rm T} dy}}
            {f_g^{h_2}(x_2,\mu^2_{\rm fact})}             
            {D_{g \rightarrow h_3}(z,\mu^2_{\rm frag})}             
        \nonumber \,.
\end{align}
The compensating subtraction term is the same expression, but now with the gluon FF replaced by its perturbative form to first order in $\alpha_s$,
\begin{equation}
D_{g \rightarrow h_3}(x,\mu^2_{\rm frag}) = \left(\frac{\alpha_s}{2\pi}\right) \log\left(\frac{\mu^2_{\rm frag}}{m^2}\right) \int_x^1 \frac{\mathrm{d}\ell}{\ell} P_{qg}\left(\frac{x}{\ell}\right) D_{Q \rightarrow h_3}(\ell)   \nonumber \,. 
\end{equation}
Consistently with our choice of scheme, also here we use the well-known zero-mass matrix elements for $d\hat{\sigma}^{gg\rightarrow  g + X}(\tau_1, \tau_2)$ with the massive expressions for $\tau_{1,2}$. The latter accounts for the fact that even if the heavy quarks do not explicitly appear in the $gg\rightarrow g+X$ process, the origins of these contributions are in diagrams where the $Q\overline{Q}$ pair is created. Without going into more details, our final expression in the GM-VFNS is eventually
\begin{equation}
\frac{d\sigma}{dP_{\rm T}dY} = \sum _{ijk}
\int \frac{dz}{z} dx_1 dx_2 
            { f_i^{h_1}(x_1,\mu^2_{\rm fact}) }
            { \frac{d\hat{\sigma}^{ij\rightarrow k}(\tau_1, \tau_2, m, \mu^2_{\rm ren}, \mu^2_{\rm fact}, \mu^2_{\rm frag})}{dp_{\rm T} dy} }
            { f_j^{h_2}(x_2,\mu^2_{\rm fact}) }            
            { D_{k \rightarrow h_3}(z,\mu^2_{\rm frag}) }             
        \nonumber \,,
\end{equation}
where the sum runs over all parton flavours and the fragmentation function is also scale dependent. Towards $p_{\rm T} \rightarrow 0$ the partonic cross sections tend to FFNS ones, but in the $p_{\rm T} \rightarrow \infty$ limit to the zero-mass $\overline{\rm MS}$ expressions. In our numerical implementation, we have taken the light-parton$\rightarrow Q$ expressions up to $\mathcal{O}(\alpha_s^3)$ from the MNR code \cite{Mangano:1991jk}, and all the remaining processes from the INCNLO code \cite{Aversa:1988vb}, up to $\mathcal{O}(\alpha_s^3)$ as well.

\vspace{-0.2cm}
\begin{figure}[htb!]
\centering
\includegraphics[width=0.45\linewidth]{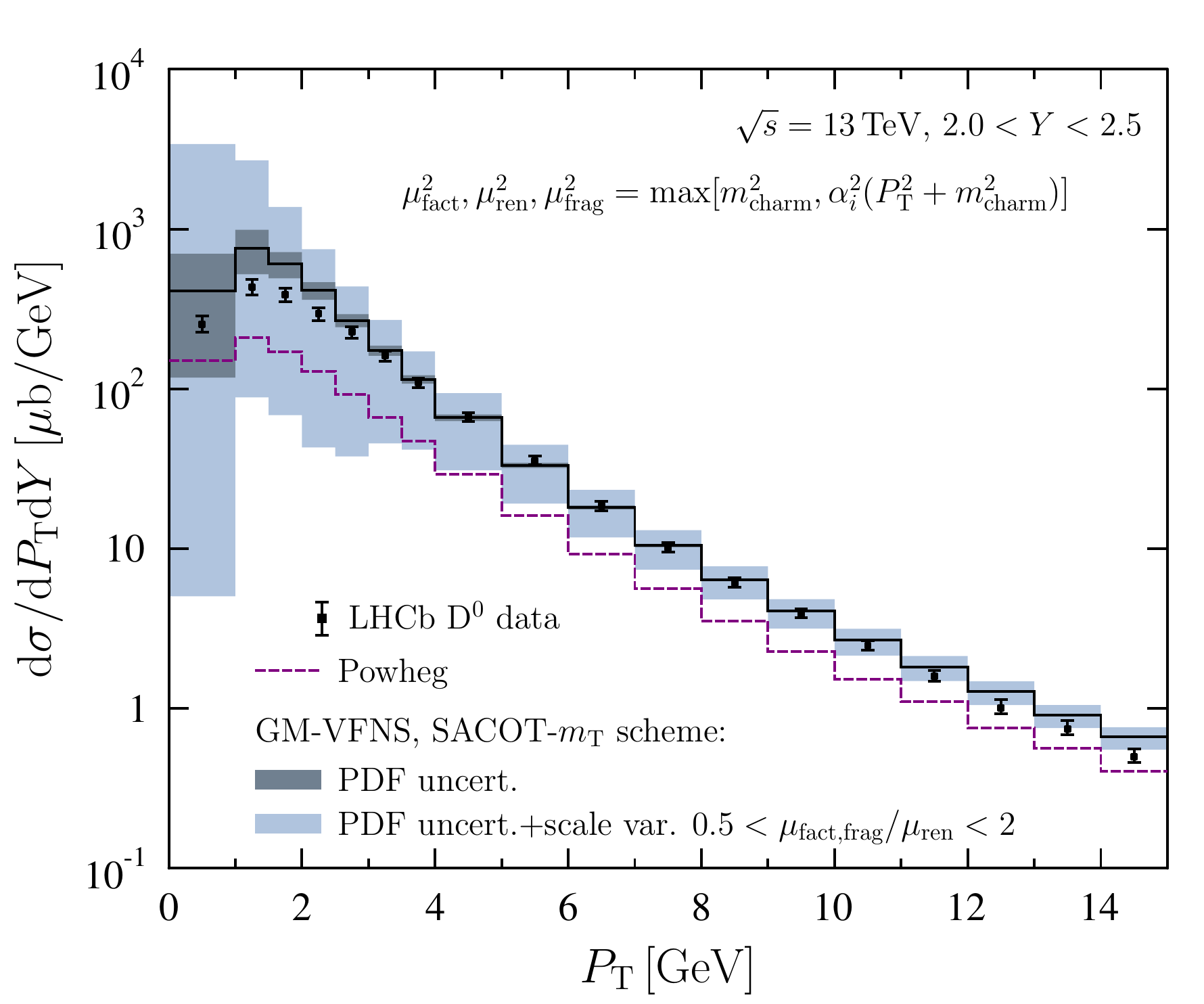}
\includegraphics[width=0.45\linewidth]{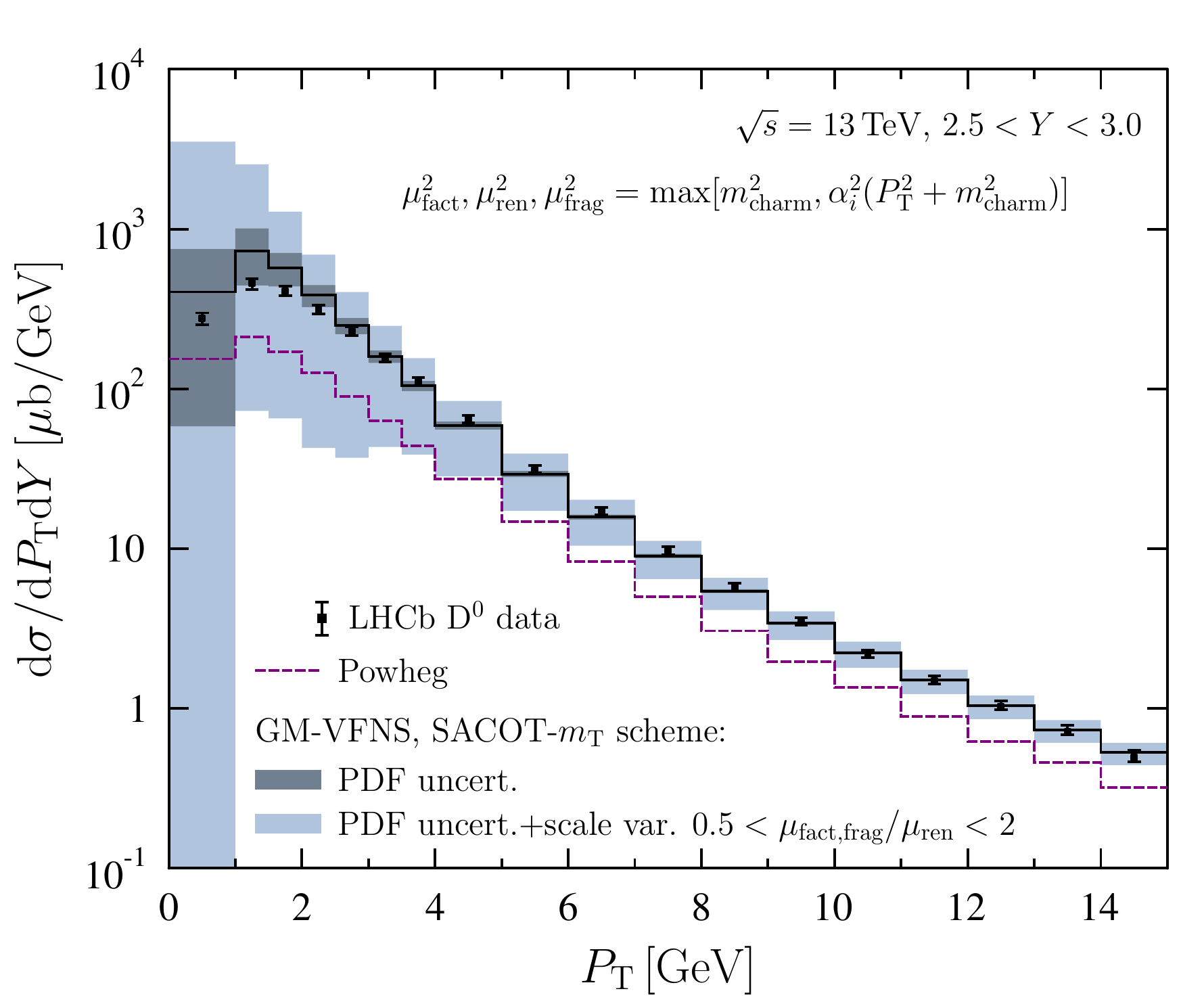}
\caption{LHCb D$^0$ data \cite{Aaij:2015bpa} in proton-proton collisions compared with our GM-VFNS calculation and \textsc{powheg}+\textsc{pythia} framework.}
\label{fig:LHCb}
\end{figure} 

\vspace{-0.2cm}
\section{Results and discussion}

\vspace{-0.2cm}
Figure \ref{fig:LHCb} presents a comparison between the LHCb 13\,TeV proton-proton data on D$^0$ mesons and our GM-VFNS theory calculation. The PDF uncertainty from NNPDF3.1 (pch) \cite{Ball:2017nwa} is shown in darker colour and the combined scale+PDF uncertainties in light blue. The FFs used are those of Ref.~\cite{Kneesch:2007ey}. The agreement is quite excellent though the scale uncertainties are large at small $P_{\rm T}$. We also compare to an approach in which the partonic $c\overline{c}$ events from \textsc{powheg} event generator \cite{Frixione:2007nw} are showered and hadronized with \textsc{pythia} 8 \cite{Sjostrand:2014zea}. Similarly to the FFNS calculations discussed earlier, the \textsc{powheg}+\textsc{pythia} setup tends to underpredict the experimental results by a factor of two. We believe the most significant reason for this is that by starting with $c\overline{c}$ pairs generated by \textsc{powheg} one misses the contributions in which the $c\overline{c}$ pair is created only later in the parton shower. Contributions like these are resummed in GM-VFNS to the scale-dependent FFs and, at high $P_{\rm T}$, e.g. the gluon-to-D contribution is around 50\% of the total cross section. In comparison to FFNS, we have also found that these contributions significantly alter the regions where the PDFs are sampled. Therefore, the use of FFNS-based calculations when fitting D-meson data with PDFs poses a potential bias.
\vspace{-0.2cm}

\section*{Acknowledgments}
\vspace{-0.2cm}

We acknowledge the Academy of Finland, Projects 297058 and 308301, as well as the Carl Zeiss Foundation and the state of Baden-W\"urttemberg through bwHPC, for support. The Finnish IT Center for Science (CSC) has provided computational resources for this work.

\end{document}